\documentclass[reprint, amsmath, amssymb, aps]{revtex4-2}

\usepackage{graphicx}
\usepackage{dcolumn}
\usepackage{bm}

\usepackage[english]{babel}
\usepackage{natbib}
\usepackage{dcolumn}
\usepackage[version=4]{mhchem}
\usepackage{bm}
\usepackage[utf8x]{inputenc}
\usepackage[T1]{fontenc}
\usepackage{array}
\usepackage{amsmath}
\usepackage{graphicx}
\graphicspath{{EPS/}}
\usepackage{float}
\usepackage{xcolor}
\usepackage{hyperref}
\usepackage{tikz,xcolor,hyperref}

\hypersetup{pdfstartview=FitH, colorlinks=true, linkcolor=blue, filecolor=blue, urlcolor=blue}

\definecolor{lime}{HTML}{A6CE39}
\DeclareRobustCommand{\orcidicon}{%
	\begin{tikzpicture}
	\draw[lime, fill=lime] (0,0) 
	circle [radius=0.16] 
	node[white] {{\fontfamily{qag}\selectfont \tiny ID}};
	\draw[white, fill=white] (-0.0625,0.095) 
	circle [radius=0.007];
	\end{tikzpicture}
	\hspace{-2mm}
}

\foreach \x in {A, ..., Z}{%
	\expandafter\xdef\csname orcid\x\endcsname{\noexpand\href{https://orcid.org/\csname orcidauthor\x\endcsname}{\noexpand\orcidicon}}
}

\newcounter{ourcount}
\begin{document}

	\title{Surfatron acceleration of the protons to high-energy in the relativistic jets} 
	
	\author{Ya. N. Istomin \orcidB{}}
	\email{istomin@lpi.ru}
	
	\affiliation{I. E. Tamm Division of Theoretical Physics, P. N. Lebedev Physical Institute of the RAS, Leninskiy Prospekt 53, Moscow 119991, Russia}
	
	\author{A. A. Gunya \orcidA{}}
	\email{aagunya@lebedev.ru}
	
	\affiliation{I. E. Tamm Division of Theoretical Physics, P. N. Lebedev Physical Institute of the RAS, Leninskiy Prospekt 53, Moscow 119991, Russia}
	
	\date{\today}
 
\begin{abstract}

    This paper describes the acceleration of high energies protons captured by electrostatic waves in the frame of jet arising in the area of instability of relativistic jet, where spiral structures are excited. The wave has a spatially heterogeneous structure of $\exp(i k_\parallel z +i m_\phi \phi)$. Protons can be captured in potential wells created by spiral waves, and thereby experience acceleration with a mechanism known as surfatron acceleration. Expressions of the maximum energy ($ E_p \simeq 10^{19} eV $) and the energy spectrum from jet parameters are obtained.
	
\end{abstract}
	

\maketitle

\section{Introduction}\label{section1}

    Axisymmetric collimated quasi-stationary ejections, called relativistic jets, arise in the process of plasma accretion onto the central black hole from the surrounding disk. Such a flow of relativistic plasma with a nonthermal nature of radiation is typical mainly for a number of active galactic nuclei (AGNs) \cite{2019ARA&A..57..467B} and microquasars \cite{2006MNRAS.370..399B}. The pioneering works, \cite{1977MNRAS.179..433B} and \cite{1982MNRAS.199..883B}, describe the prerequisites for the emergence of axisymmetric jets. The quasi-cylindrical structure of the jet is formed by the ratio of the prevailing toroidal magnetic field over the longitudinal one. In the interaction of a magnetic field with an unstable plasma flow arising due to the Kelvin-Helmholtz instability \cite{1984ApJ...287..523H}, \cite{1986ApJ...303..111H} in the layer between the external reverse and internal direct electric jet current, in the latter, spiral structures are formed that have surfaces with potential wells. A particle entering the region of an electrostatic well experiences multiple reflections from its walls. There is a kind of capture of a particle by an electrostatic wave. In this case, the particles, on average, move along with the wavefront. When the wave moves, the trapped particles, oscillating, gain energy due to the energy of the wave. Here, formally, the particle is accelerated by the Lorentz force. This is the process of surfatron acceleration. Further, when the wave attenuates and the blocking potential decreases, the accelerated particle leaves the region of the electrostatic wave.

    In this paper, the authors consider high-energy protons as accelerated particles, which also have the ability to be accelerated in the black hole magnetosphere and in the jet to significant energies by regular acceleration mechanisms considered earlier in the papers \cite{2009Ap&SS.321...57I}, \cite {2020MNRAS.492.4884I} and \cite{PhysRevD.102.043010}. In the pioneering work, \cite{1983PhRvL..51..392K}, where the surfatron acceleration mechanism was proposed, laboratory plasma was considered, where electrons were used as accelerated particles. In our case of astrophysical plasma, however, the acceleration of heavy particles (protons), which are also captured by the electric field of the waves, takes place predominantly. The absence of synchrotron radiation, i.e. the absence of the most powerful radiative energy loss channel affecting electrons, contributes primarily to the acceleration of protons, which is reflected, in particular, in the fact that cosmic rays mainly consist of high-energy protons, up to superhigh energies.
    
    The appearance of electrostatic waves in the jet, into which charged particles are trapped and accelerated by the surfatron, is the result of instability that occurs at the periphery of the jet in the region where the reverse electric current passes. In the central part of the jet, the current flows along the jet axis in one direction, while to close it, the current must flow in the opposite direction at the periphery. As a result of the interaction of oppositely directed currents, the equilibrium of the plasma flow is disturbed. In addition to stationary magnetic fields: longitudinal (poloidal) $B_z$ and toroidal magnetic field $B_\phi$ and fluxes: longitudinal velocity $v_z$ and rotation $v_\phi$, waves arise that have a spatial dependence in the form of first harmonic, ${ \bf B}_1, {\bf v}_1 \propto \exp(ik_\parallel z+im_\phi \phi)$. The resulting waves can capture protons and, as is the case with surfatron motion, their acceleration. This is a new mechanism for the acceleration of charged particles in astrophysical plasma.
    
    Separately, it is worth noting that although the surfatron acceleration mechanism and the impact mechanism accelerate particles in a wave, the final energies achieved as a result of acceleration have a significant difference in energies. Thus, the limiting energy for the impact mechanism is up to $10^{15}$ eV versus, for example, $10^{18}$ eV given as an example in this paper for the surfatron mechanism.
    
    The problem of particle acceleration by a jet, solved in this paper, is based on the results of \cite{2018MNRAS.476L..25I}, where the MHD solution and the mechanism of wave generation in the jet itself are considered in detail.
    
    The structure of the electromagnetic fields of the jet necessary for surfatron acceleration is described in the section (\ref{section2}). The equations of motion and their solution are presented in the section (\ref{section3}). Consideration of specific jets in AGN is done in the section (\ref{section4}). A discussion of the results and their interpretation in the context of the physical parameters of the AGN is presented in the section (\ref{section5}).

\section{Electromagnetic field structure}\label{section2}

    We are working in the frame of the stationary MHD approximation in a coordinate system moving with the plasma flow along its axis with the velocity $v_z$.
\begin{eqnarray}\label{fieldsMHD}
    &&\nabla(n{\bf v})=0; \\ \nonumber
    &&({\bf v}\nabla){\bf v}=-\frac{1}{n}\nabla P(n)-\frac{1}{8\pi n}\nabla B^2+ \frac{1}{4\pi n}({\bf B}\nabla) 
{\bf B}; \\ \nonumber
    &&\mathrm{curl}[{\bf vB}] =0; \\ \nonumber
    &&\nabla{\bf B}=0. \nonumber
\end{eqnarray}
    Here $ n $ is the plasma density of the jet, ${\bf v}$ is the plasma velocity, ${\bf B}$ is the magnetic field, $P(n)$ is the plasma pressure, which is a function of the density through the equation of state of the matter in the jet.

    Suppose first that all quantities depend only on the cylindrical radius coordinate $r$, that is, the jet is axisymmetric and homogeneous along axis $z$. In this case, the velocity and the magnetic field can be expressed as functions of $r$,
$$
    {\bf v}={\bf e}_\phi v_{\phi}(r), \, {\bf B} ={\bf e}_\phi B_\phi(r)+{\bf e}_z B_z(r).
$$
    Velocity $v_\phi$ is the speed of the rotation of the jet, $B_z$ is the longitudinal magnetic field and $B_\phi$ is the toroidal magnetic field. It should be noted that the radial velocity $v_r$ and the radial magnetic field $B_r$ in the case of only cylindrical radius dependence are zero by virtue of condition $\nabla(n {\bf v})=0, \, \nabla {\bf B}=0$ (\ref{fieldsMHD}) and finiteness condition at $r=0$. The perturbed components should satisfy the next set of expressions:

\begin{eqnarray}
    &&\nabla{\textbf n_1}v + \nabla{\textbf v_1}n = 0, \\ \nonumber
    &&\nabla{\textbf B_1} = 0, \\
    &&({\textbf B}\nabla){\textbf{v}_1} + ({\textbf B_1}\nabla){\textbf{v}} + {\textbf B}\nabla{\textbf{v}_1} - {\textbf B}(\nabla{\textbf{v}_1}) - {\textbf B_1}(\nabla{\textbf{v}}) = 0.\nonumber
\end{eqnarray}

    As shown earlier in \cite{2018MNRAS.476L..25I}, such a solution does not satisfy the jet equilibrium condition in the radial direction $r$ for all radius values. In the reverse current region, closer to the jet periphery, spatially inhomogeneous perturbations should arise, which can be represented as a superposition of harmonics along $z$ and $\phi$ coordinates
\begin{eqnarray}
    &&n=n(r)+n_1(r)\exp\{ik_\parallel z+im_\phi \phi\}, \\ \nonumber
    &&P=P(r)+P_1(r)\exp\{ik_\parallel z+im_\phi \phi\}, \\ \nonumber
    &&{\bf v}={\bf e}_\phi v_\phi(r)+{\bf v}_1(r)\exp\{ik_\parallel z+im_\phi \phi\}, \\ \nonumber
    &&{\bf B}={\bf e}_\phi B_\phi(r)+{\bf e}_z B_z(r)+{\bf B}_1(r)\exp\{ik_\parallel z+im_\phi \phi\}, \\ \nonumber
    &&{\bf E}={\bf e}_r E_r(r) + E_1(r) \exp\{ik_\parallel z+im_\phi \phi\}.
\end{eqnarray}
    Here $k_\parallel$ and $m_\phi $ are the longitudinal and azimuthal wave numbers, respectively. 

    The perturbed electric field $E_1$ arises against the background of the regular radial field $E_r$ and, by definition, cannot exceed it because it corresponds to the maximum energy generated by the black hole, extracted by the Blandford-Znaek mechanism \cite{1977MNRAS.179..433B}. As in the case of proton acceleration in a jet by a regular electromagnetic field \cite{PhysRevD.102.043010}, the amplitude of the radial regular electric field $E_r$ is defined as the voltage generated by the black hole $U = B_p \cdot r_g^2 \cdot \Omega_H / 2c $ [eV]. Here, $B_p$ is the equipartition magnetic field in the vicinity of the black hole, $r_g$ is the gravitational radius of the black hole, and $\Omega_H$ is the angular velocity of the black hole. As in earlier papers \cite{PhysRevD.102.043010}, \cite{2020MNRAS.492.4884I}, it is assumed that the energy of this field is the maximum for a system with a black hole. Since the amplitude of the perturbed electric field has a lower energy, the energy of the particle is also limited by the value $E_1 < E_r$. Moreover, the superiority of the regular field also plays a key role because, as was shown in \cite{PhysRevD.102.043010}, it acts in the direction of the particle leaving the relativistic jet region, which also determines the motion of the particle from the wave region potential well on the jet surface. Proceeding from this, the amplitude of the perturbed electric field can be estimated as $E_1 \simeq (\xi / R)\cdot E_r$.
    
\begin{eqnarray}\label{d1}
    &&E_1 = -\frac{1}{c}([\vec v_1 \times \vec B] + [\vec v \times \vec B_1]), \\
    &&E_r = -\frac{1}{c} v_\phi B_z.
\end{eqnarray}

    First, we describe the perturbed component of electric field ${\bf E}_1$. The first term in (\ref{d1}) for the electric field has the unperturbed longitudinal $B_z$ and azimuthal $B_\phi$ components of the magnetic field and the perturbed velocities $v_{1r}$, $v_{1\phi}$, $v_ {1z}$. The second term for the electric field has all the perturbed magnetic field components $B_{1r}$, $B_{1\phi}$ and $B_{1z}$ and the unperturbed toroidal velocity component $v_\phi$. The values with index 1 in the expression (\ref{d1}) are perturbed quantities. Quantities without an index are unperturbed.

    According to \cite{2018MNRAS.476L..25I}, the values of the perturbed field and velocity components are as follows:

\begin{eqnarray}\label{d5}
    &&v_{1r} = -i \frac{\xi}{R}\frac{m_{\phi} v^{'}_\phi}{B^{'}_z} B_{\text{1z}}, \\ \nonumber  
    &&v_{1\phi} = \frac{v^{'}_\phi}{B^{'}_z} \frac{R B^{'}_z k_\parallel+2 m_{\phi} B^{'}_\phi}{R B^{'}_z k_\parallel+m_{\phi} B^{'}_\phi} B_{\text{1z}}, \\ \nonumber 
    &&v_{1z} = \frac{m_{\phi} v^{'}_\phi}{R B^{'}_z k_\parallel+m_{\phi} B^{'}_\phi} B_{\text{1z}}, \\
    &&B_{1r} = -i\frac{\xi}{R} \frac{k_\parallel R B^{'}_z +m_{\phi} B^{'}_\phi}{B^{'}_z} B_{\text{1z}}, \ B_{1\phi} =\frac{{B^{'}_\phi}}{{B^{'}_z}} B_{\text{1z}}. \nonumber 
\end{eqnarray}
    Here, the expression $\exp\{ik_\parallel z+im_\phi \phi\}$ is temporarily omitted. It should be noted that the magnitudes of the longitudinal $B_z$ and toroidal $B_\phi$ magnetic fields, as well as the toroidal velocity component $v_\phi$ tend to zero at the jet periphery. Taking this into account, we can represent such quantities near the jet boundary through their values that depend linearly only on the radius: $B_z = \xi B^{'}_z$, $B_\phi = \xi B^{'}_\phi$, $ v_\phi = \xi v^{'}_\phi$. Here, the distance $\xi = R - r$ is measured from the edge of the jet. Then, considering (\ref{d5}), we obtain expressions for the components of the electric field:
    
\begin{eqnarray}\label{d7}    
    &&E_{1r} = \frac{2}{c} v_{\phi } B_{\text{1z}}\exp\{ik_\parallel z+im_\phi \phi\}, \\ \nonumber
    &&E_{1\phi} = \frac{i \xi m_{\phi } }{cR} v_{\phi } B_{\text{1z}}\exp\{ik_\parallel z+im_\phi \phi\}, \\
    &&E_{1z} = \frac{i\xi  k_\parallel }{c} v_{\phi } B_{\text{1z}}\exp\{ik_\parallel z+im_\phi \phi\}. \nonumber
\end{eqnarray}
    It is not difficult to see that the electric field ${\bf E}_1$ is potential
\begin{equation} \label{psi}
    {\bf E}_1 =-\nabla\Psi, \, \Psi=-\frac{\xi }{c} v_\phi B_{1z}\exp\{ik_\parallel z+im_\phi \phi\}.  
\end{equation}

\begin{figure}
    \centering
    \includegraphics[width=0.45\textwidth]{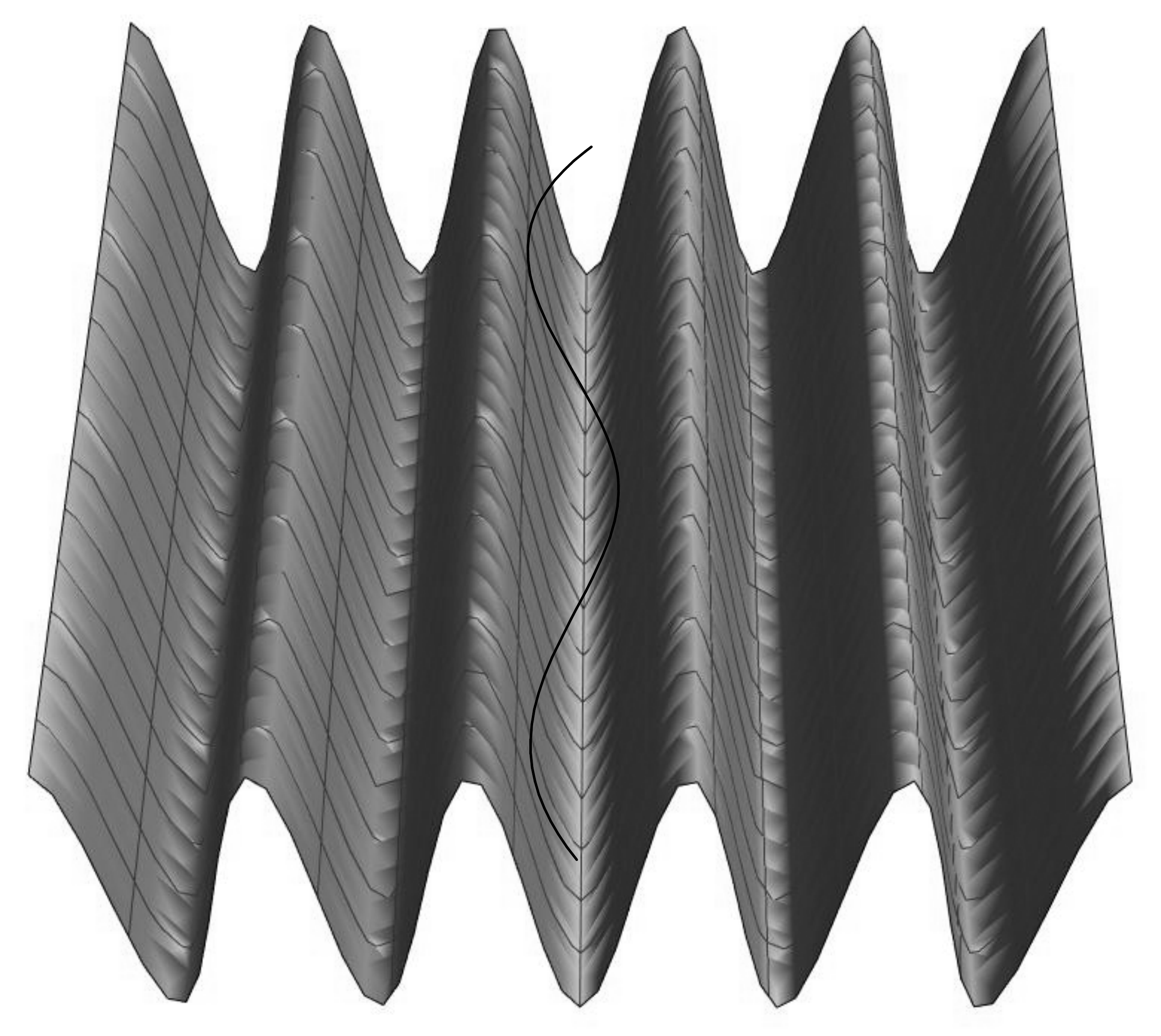}
    \caption{3D representation of a spiral wave with $k_\parallel R=3$, $m_\phi =5$. The black line shows the trajectory of the captured proton. Bends are reflections from the humps of the potential}
    \label{p1}
\end{figure}

    Here, we are looking for a steady, stationary, though not stationary, solution for the electromagnetic fields in the jet. Thus, at the periphery of the jet, waves appear that have a wave spiral structure, which is necessary for the appearance of the effect of surfatron acceleration of particles. The waveform, with $k_\parallel R=3$, $m_\phi=5$ is shown in the figure (\ref{p1}).

\section{Surfatron acceleration}\label{section3}

    Particles moving near the wells of the $\Psi$ potential can be trapped in them. The phase of the wave seen by such particles will be almost constant along the line $z=-(m_\phi/k_\parallel)\phi$. For them, the speed will be almost constant, $u_z= -(m_\phi/k_\parallel R)u_\phi$. The average Lorentz force $e[\bf u B]/c$ acting on them will be constant on average and will lead to the acceleration of trapped particles in the radial direction
\begin{equation}
    \frac{d p_r}{dt}=\frac{e u_\phi B_z}{c}\left(1+\frac{m_\phi}{k_\parallel R}\frac{B{'}_\phi}{B^{'}_z}\right).
\end{equation}
    Thus, the Lorentz factor of particles will increase $\gamma$,
\begin{equation}\label{gamma} 
    \frac{d\gamma}{dt}=\alpha\omega_c\frac{u_\phi}{c}, \, \alpha=\left(1+\frac{m_\phi}{k_\parallel R}\frac{B_\phi}{B_z}\right). 
\end{equation}
    Here the quantity $\omega_c$ is the non-relativistic cyclotron frequency of the proton in the magnetic field $B_z, \, \omega_c=eB_z/mc$. We believe that the particle has accelerated to relativistic energies ($u_r \simeq c$). The coefficient $\alpha$ generally varies along the jet radius $r$, $|\alpha|\simeq 1$ in the inner part of the jet and is equal to $|\alpha|\sim 10$ on the periphery. To accelerate particles in the positive direction $r$ ($u_r>0$), from the center to the periphery of the jet, the value of $\alpha$ must be positive. Since the amplitude of the perturbed field $B_1$ is maximum at the periphery of the jet, the acceleration of protons is most effective when it moves outward.

    It should be noted that if electrons could also be captured into spiral waves, then the surfatron mechanism would lead to the same values of their energies as protons. However, the resulting radiation during the relativistic motion of a charged particle in a periodic potential, called undulator radiation, will cause strong energy losses for electrons. In this way. just as in the case of synchrotron radiation, high electron energies, such as those of protons, are far from being reached.

    However, with surfatron acceleration, with an increase in the Lorentz factor of the particle $\gamma$, the inertia force acting on the particle in the direction of its capture into the well increases $\Psi$ (\ref{psi}), $F_i=dp_\phi/dt$
\begin{equation}
    F_i = mu_\phi\frac{d\gamma}{dt}=\alpha m\omega_c c\frac{u_\phi^2}{c^2}.
\end{equation}
    Under the condition that the inertia force $F_i$ exceeds the "capture" force $|eE_\phi|$ (\ref{d7}), the particle escapes capture, and the surfatron acceleration ceases to operate. This condition determines the region of acceleration: if the capture of the particle occurred at the point $r=r_i, \, \xi=R-r_i=\xi_i$, then the exit from the capture, i.e. acceleration will stop at the point $\xi=\xi_f$, $\xi_i>\xi_f$
\begin{equation}
    \xi_f^2=\frac{\alpha}{m_\phi}\left(\frac{u_\phi}{c}\right)^2\frac{B_z}{B_1z}\frac{Rc}{\Omega}.
\end{equation}
    Here we denote the quantity $u^{'}_\phi$ $ (u_\phi=u^{'}_\phi\xi)$ as $\Omega$. This value is equal in the order of magnitude to the angular velocity of rotation of the inner part of the jet. Keeping in mind that $d\gamma/dt = -c d\gamma/d\xi$, the integral of the equation (\ref{gamma}) gives
\begin{equation} \label{max}
    \gamma=\gamma_i-\alpha\frac{\omega_c u_\phi}{2c^2}(\xi_f^2-\xi_i^2).
\end{equation}
    Since the maximum value of $\xi$ is equal to $R$, the condition for surfatron acceleration in a spiral jet has the form $\xi_f<R$. This inequality gives a constraint on the toroidal velocity of accelerating particles
\begin{equation}
    \frac{u_\phi}{c}<\left(\frac{m_\phi B_{1z}}{\alpha B_z}\frac{\Omega R}{c}\right)^{1/2}.
\end{equation}
    Since the ratio $\Omega R/c>>1$, this restriction is not essential. The value of the maximum Lorentz factor of the particle $\gamma_m$, followed from the equation (\ref{max}), achieved by the surfatron acceleration mechanism
\begin{equation}
    \gamma_m\simeq \alpha\frac{\omega_c R u_\phi}{2c^2}
\end{equation}

    The evaluation of the finite energy of particles directly depends on the range of the jet parameters. As an instance, it might be quite interesting to consider the parameters of the M87 jet, which is also discussed in \cite{PhysRevD.102.043010}. Here, as it was mentioned before, $\alpha \simeq 10$. This coefficient depends on relation $B_\phi / B_z $, which for M87 is about $<B_\phi  / B_z>\simeq 8...10 $. The relation $m_\phi/k_{II}R \simeq 5 / 3$ \cite{2018MNRAS.476L..25I}. The cyclotron frequency $B_z(e/mc)$, where $B_z \simeq 10^{-1}$ G for jet base and about $10^{-3}$ G for potential well area with jet radius R $\geq 10^{19}$ cm. It’s interesting to note that this set of parameters with toroidal particle velocity $u_\phi$ has the order of Hillas criterion \cite{1984ARA&A..22..425H}.

    If $\omega_c\sim 10 s^{-1}$, then the maximum factor is $\gamma_m \sim 10^9$. This is in the laboratory coordinate system, where the jet moves along the axis with a speed of $v_z$. If the Lorentz factor of this motion is equal to $\Gamma\sim 10$, then the maximum attainable energy by surfatron mechanism will be greater and for M87 surfatron accelerated protons be equal $E = \Gamma \gamma_m m_p \sim 10^{19}$ [eV]. The maximum energy which is achieved by regular electric field accelerated protons, according \cite{PhysRevD.102.043010} is $E \sim 10^{20}$ [eV]. So, eventually, the energy of surfatron accelerated protons could consist about $10\%$ of total energy generated by the supermassive black hole of M87 into the relativistic jet.

    An interesting question is about the energy spectrum of accelerated particles $\gamma>>1$. Assuming the plasma density in the jet to be more or less constant in the jet cross-section, we have $dN/dr \propto r$. The quantity $N$ is the density of accelerated protons. Further, $dN/d\xi\propto -(R-\xi)$, and for high-energy particles $d\gamma/d\xi=(\alpha\omega^{'}_c u_\phi/c^2 )\xi$, we get $dN/d\gamma\propto -(c^2/\alpha\omega^{'}_c u_\phi(R/\xi-1)$. Since $\xi\propto \gamma ^{1/2}$, then the resulting spectrum has the form
\begin{equation}
    \frac{dN}{d\gamma} \propto -\left[\left(\frac{\gamma_m}{\gamma}\right)^{1/2}-1\right].
\end{equation}
    It can be seen that the spectrum is hard, $dN/d\gamma\propto \gamma^{-1/2}$ at high energies and cuts off at $\gamma =\gamma_m$.

\section{Spiral jet}\label{section4}

\begin{figure*}
    \includegraphics[width=0.45\textwidth]{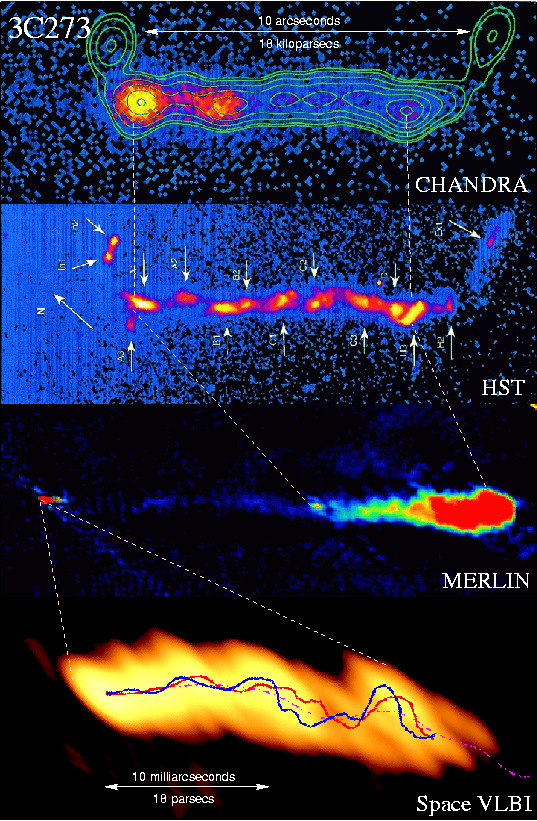}
    \includegraphics[width=0.45\textwidth]{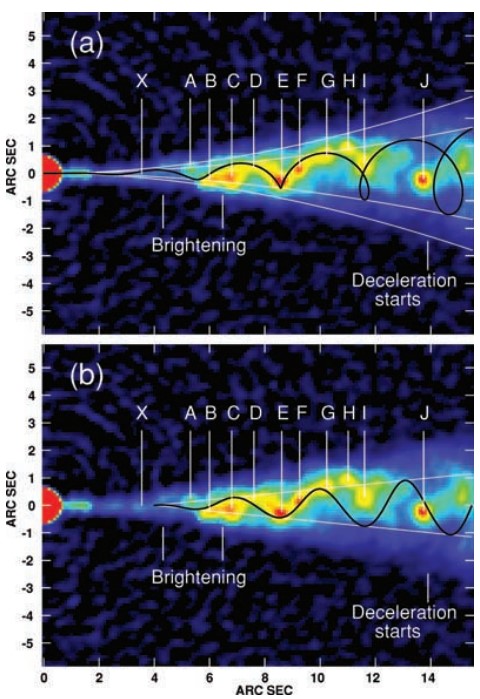}
    \caption{Base structures of jets NGC315 (left) \cite{2007MNRAS.380....2W} and 3C273 (right) \cite{2001Sci...294..128L}}
    \label{p3}
\end{figure*}

    The most accurate resolution of the image of the relativistic jet M87 \cite{2021ApJ...923L...5P}) at frequencies of 4-18 GHz clearly indicates the regular helicity of the structure, which is especially noticeable in the region of the conical jet at a distance of 340 pc from the nucleus, which also quite suggests the development of such a structure directly at the base of the jet near the nucleus, which, for example, was found at the base of the 3C273 (fig. \ref{p3}) jet in \cite{2001Sci...294..128L}. A similar picture was also observed in the inner jet of the radio galaxy NGC315 (fig. \ref{p3}) \cite{2007MNRAS.380....2W}. All this gives reason to believe that helicity can develop from the base and further into the region of the conical jet. It is interesting to consider precisely the conical region of the jet, and as the most striking example, we will further consider the M87 jet.

\begin{figure*}
    \includegraphics[width=0.45\textwidth]{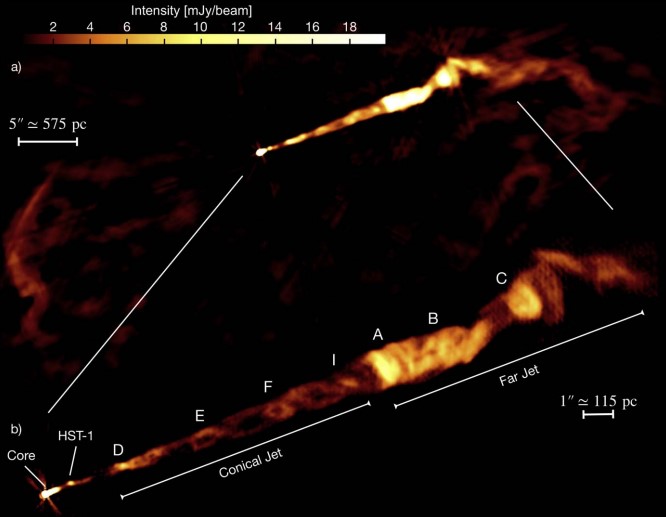}
    \includegraphics[width=0.45\textwidth]{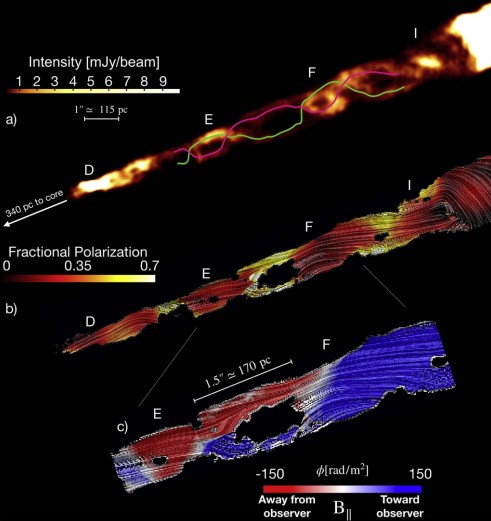}
    \caption{Morphological and polarization interpretation of the M87 relativistic jet base image \cite{2021ApJ...923L...5P}};
    \label{p2}
\end{figure*}

    The striking pattern of polarization in the M87 image (fig. \ref{p2}) illustrates a periodic sequence of the brightest and faintest nodes of a $\sim$ 170 pc spiral jet for an outside observer. The electrostatic waves arising in the spirals can be filled with sufficiently dense plasma up to a value of the order of the average jet plasma density $n \simeq 10^{-1} cm^{-3}$. In this case, it is possible to capture a significant number of particles and their acceleration. However, in reality, to determine the number of particles trapped in potential wells, it is necessary to determine the known low probability of their capture, which is a separate, rather complicated problem that is not considered in this paper.

    Particles trapped in helical electrostatic wells are accelerated by the surfatron mechanism as described above. Of these, the most interesting are protons that have traveled the maximum possible radial distance while being trapped. The exit of the proton from the potential well occurs due to the forces of inertia due to their acceleration when the force of capture into the well is exceeded.

    Registration of protons accelerated by just such a mechanism in a jet and their separation from other high-energy protons is currently extremely difficult due to the insufficient resolution of neutrino telescopes, as well as gamma-ray telescopes. Moreover, the concentration of matter in the potential wells is significantly lower than in the jet. Natural radiative dissipation in the form of undulator radiation, which qualitatively replaces synchrotron radiation in such mechanisms, is not significant for protons and cannot be considered as a reliable indicator of the proton acceleration process.

    The time of the proton's movement along the jet spiral determines the probability of collision with photons or other protons. A particle trapped on the periphery of the jet in the $\xi$ region travels a significant distance along a complex trajectory along the jet axis $z$, and, therefore, has a high probability of colliding with other particles also because the trajectory is complicated by oscillations in the potential well itself when approaching the humps of the potential. Such collisions should be present to a greater extent on the periphery of the jet. Since gamma photons are less likely to reach the observer due to interaction with the intergalactic medium, it is more interesting to consider neutrinos.

    The expected neutrino energies based on the energy of colliding protons are of the order of $E_\nu = 0.05E_p$, i.e. $\sim 10^{16}$ eV. The detection of high-energy neutrinos at the base of relativistic jets has previously taken place \cite{2020ApJ...894..101P}. The maximum currently registered neutrino energy is 290 TeV of the IceCube-170922A event from the blazar TXS 0506+056 \cite{2018Sci...361..147I}. The occurrence of such neutrino events, for example, in M87, which can most accurately be associated with surfatron accelerated protons, is most likely to be detected in the conical jet region (between the \textit{D} and \textit{I} (fig. \ref{p2}) regions). The reason is that the density of neutrino events in the region of the nucleus is much higher and may have a mixed origin. Moreover, the difference in the number of events and energies between $E_\nu$ from protons with energies up to 10-100 TeV in the medium of a conical jet and surfatron protons is quite large, which also more unambiguously allows us to draw a conclusion about the surfatron nature of the acceleration of primary neutrino-forming protons. The number of protons with the maximum energy obtained as a result of surfatron acceleration can be determined by the balance ratio of the energies of matter and field $n_p < B^2/(4\pi \gamma_p m_p c^2) = 10^{-13}$ cm$^{ -3}$. Here, the field $B\simeq10^{-3}$ G is assumed to be homogeneous, although in the spiral structure between the nodes the toroidal $B_\phi$ and longitudinal $B_z$ components of the jet magnetic field alternately dominate.

    There are two main reaction channels proton-proton and photon-proton collisions between, for example, a surfatron-accelerated proton and an SSC photon with frequencies up to $10^{15}$ Hz.

    When passing from a parabolic to a conical region, the jet plasma concentration decreases with radius as $n \propto r^{-2}$. Therefore, the number of reactions is much less in the conical region of the jet compared to the core. Among the main scenarios, the proton-proton $pp$ mechanism seems unlikely due to a more rarefied medium than, for example, near the nucleus. The interaction time of high-energy protons accelerating in a spiral jet with protons with a concentration of the order of $n = 10^{-1}$ cm$^{-3}$ in a magnetic field of $10^{-4}$ G is rather long,
    
\begin{equation}    
    t_{pp} \simeq \frac{1}{c n \sigma_{pp}} \simeq 10^{14} \, s
\end{equation}    

    where $\sigma_{pp} \simeq 10^{-26}$ cm$^{2}$ is the $pp$ interaction cross-section. It is important to note here that the cross-section increases with increasing proton energy \cite{2004vhec.book.....A}, i.e. $\sigma_{pp} (E_p) = 30[0.95+0.06ln(E_{kin}/1 GeV)] mb$, where $E_{kin} = E_p - E_{rest} \simeq E_p$ and for $E_p ^{max} \simeq 10^{19}$ eV will be 70 mb, which, however, will not significantly change the picture.

    More interesting is the process of $p\gamma$ interaction, with the possible formation of both neutrinos and gamma photons. For the most effective interaction with SSC photons with a minimum energy of $\epsilon_{ssc} > 10$ eV, protons should be accelerated to $E_p \simeq 300$ MeV $/ \epsilon_{ssc} m_p \simeq 3\cdot 10^{17} $ eV. The SSC photon density at the jet periphery can be defined as $n_{ssc} = L_{ssc}/(4 \pi R_j^2 \epsilon_{ssc}) \, cm^{-3}$. For the jet conical region $R_j \simeq 10^{19}$ cm, the luminosity of SSC photons is $L\simeq10^{41}$ erg $\cdot$s$^{-1}$ \cite{2012A&A...544A..96A}, then the concentration is $n_{ssc} \simeq 10$.

    For an average photopion production cross-section of $10^{-28}$ cm$^2$, the proton cooling rate can be estimated from the following formula \cite{1990ApJ...362...38B}; \cite{2015MNRAS.447...36P}, where the SSC photon energy is considered in the reference frames of the observer and proton as $\epsilon = 2 \gamma_p \epsilon_{ssc}$, $\epsilon_{ssc}$, respectively. Moreover, the threshold energy corresponds to $\epsilon_{th}$ = 145 MeV.

\begin{equation} 
    t_{p\gamma}^{-1}(\gamma_p) = \frac{c}{2\gamma_p^2} \int_{\epsilon_{th}}^{\infty} \epsilon \sigma_{p\gamma} \,d\epsilon \int_{\epsilon_{s}}^{\infty} \frac{n_{ssc}(\epsilon_{ssc})}{\epsilon_{ssc}^2} \,d\epsilon_{ssc}
\end{equation} 

    According to the formula above, the time is $\simeq 4 \cdot 10^9 \, s$. This estimate already illustrates a much shorter cooling time, which means a more efficient process of neutrino production than with $pp$ interactions. Thus, the rate of photomeson processes in AGN jets is very sensitive to proton energy, namely, it requires $E_p > 10^{16}$ eV, which is obvious and achievable with surfatron acceleration in regions far from the nucleus and in the parabolic region of the jet. A significant advantage of proton acceleration by such a mechanism is that, due to the relatively long acceleration distance, the proton can regain its energy lost during photomeson cooling.

    Based on the calculated cooling time, the cooling distance is $l_{p\gamma} = c t_{p\gamma} \simeq 10^{20}$ cm. Note that the length of the spiral trajectory along the conical part of the jet is greater than the length along the $z$ axis $l_{jet} < l_{sp}$. Here, based on the figure (\ref{p2}) $l_{sp} \simeq 3 l_{jet}$. Acceleration time $t_{acc} = \Gamma m_p c \gamma_p / q B_z$. Lorentz factor of the M87 jet $\Gamma \simeq 3.4$ \cite{2003NewAR..47..629L}, then, taking into account the rest of the previously mentioned parameters, $t_{acc} \simeq 3 \cdot 10^8$ s, which corresponds to $l_{acc} = c t_{acc} \simeq 10^{19}$ cm. Thus, we have $l_{acc} < l_{p\gamma} < l_{sp}$, i.e. the proton that enters the capture has time to gain maximum energy before colliding with the SSC photon because $l_{p\gamma} / l_{acc} \simeq 10$. Theoretically, this process can be repeated $l_{sp}/l_{p\gamma} > 30$ times for one proton.

    Replacing the conical shape of the jet with a cylindrical one, and also representing the spiral periphery $\xi$ thick as uniform in the concentration of high-energy protons, the expected flux, as well as the luminosity of neutrinos from surfatron protons, can be estimated based on the total flux of surfatron-accelerated protons up to high energies $3\cdot10^{ 34}$ cm$^{-2}$, which corresponds to a luminosity of $3\cdot10^{41}$ erg$\cdot$s$^{-1}$. Then the flux of surfatron neutrinos is about $\simeq 10^{39}$ erg$\cdot$s$^{-1}$. The total neutrino flux from the M87 jet can average $L_\nu = 0.05L_{jet} = 0.05 \cdot 10^{44.45} \simeq 10^{42}$ erg$\cdot$s$^{-1} $.

    In addition to neutrinos, gamma photons with energies $E_\gamma >0.2E_p \simeq 10^{18} $ eV. Such photons can also be obtained in other processes by interacting with CMB photons. The process of CMB interaction is interesting in that the interacting protons must already leave the potential well of the jet with the maximum energy to interact with the CMB photons surrounding the galaxy with the jet.

    Gamma-ray photons of one of the brightest TeV blazars 1ES 0229+200 are predicted to have energies in the range beyond 10-100 TeV, which, however, will require the full operation of the future CTA array.

\section{Conclusions}\label{section5}

    The process considered in this paper, despite its novelty in the astrophysics of active nuclei and relativistic jets, can become one of the main ones. Due to the lower concentration of the plasma of the conical jet than that of the vicinity of the nucleus, the protons accelerated by the surfatron mechanism undergo fewer collisions and, in addition, experience insignificant natural radiative losses. The area of proton emission from the jet is much larger than near the nucleus. All this also means that the surfatron acceleration of protons, in the region of a conical jet far from the galactic disk, can make a significant addition to the total flow of ultrahigh-energy protons from AGNs. This, in turn, can also significantly enhance the high-energy component of the gamma background from surfatron protons around the AGN galaxy. In addition, the bulk Lorentz factor of some quasars can reach much higher values, which means even more energy.

\newpage

\bibliography{References}

\end{document}